\documentclass[journal]{IEEEtran}
\usepackage{mathbbol}
\usepackage{amsthm,amsmath,amssymb}
\usepackage{dsfont}
\usepackage{mathrsfs}
\usepackage{amssymb,amsmath}
\usepackage{algorithmic}
\usepackage{algorithm}
\usepackage{url,subfigure,epsfig,graphicx}
\usepackage{float}
\usepackage{makecell}
\usepackage{CJKutf8}
\usepackage{stfloats}
\usepackage{graphicx}
\usepackage{subcaption}
\usepackage{color}
\ifCLASSOPTIONcompsoc

\usepackage{float}
\usepackage{makecell}
\usepackage[x11names,dvipsnames,table]{xcolor} 
\usepackage{colortbl}
\usepackage{multirow}
\usepackage{subcaption} 
\usepackage{lipsum}
\definecolor{mygray}{gray}{0.6}
\definecolor{myblue}{rgb}{0.8,0.85,1} 
\usepackage{array}
\newcolumntype{L}[1]{>{\raggedright\let\newline\\\arraybackslash\hspace{0pt}}m{#1}}
\newcolumntype{C}[1]{>{\centering\let\newline\\\arraybackslash\hspace{0pt}}m{#1}}
\newcolumntype{R}[1]{>{\raggedleft\let\newline\\\arraybackslash\hspace{0pt}}m{#1}}

\usepackage{array, tabularx, boldline}
\usepackage{eurosym}
\usepackage{amstext} 
\DeclareRobustCommand{\officialeuro}{%
	\ifmmode\expandafter\text\fi
	{\fontencoding{U}\fontfamily{eurosym}\selectfont e}}

  \usepackage[nocompress]{cite}
\else

  \usepackage{cite}
\fi

\ifCLASSINFOpdf

\else

\fi

\begin{document}

\begin{CJK}{UTF8}{gbsn}

\title{Multi-layer Digital Twin System for Future Mobile Metaverse}



\author{
\IEEEauthorblockN{Gaosheng Zhao,~\IEEEmembership{Graduate Student Member,~IEEE,} and~Dong In Kim,~\IEEEmembership{Life Fellow, IEEE}}
}

\maketitle

\begin{abstract}
In the upcoming 6G era, the communication networks are expected to face unprecedented challenges in terms of complexity and dynamics. Digital Twin (DT) technology, with its various digital capabilities, holds great potential to facilitate the transformation of the communication network from passive responding to proactive adaptation.
Thus, in this paper, we propose a multi-layer DT system that coordinates local DT, edge DT, and cloud DT for future network architecture and functions. In our vision, the proposed DT system will not only achieve real-time data-driven decision-making and digital agent functions previously handled by centralized DT, but will do so in a more distributed, mobile, layer-by-layer manner.
Moreover, it will supply essential data, pre-trained models, and open interfaces for future metaverse applications, enabling creators and users to efficiently develop and experience metaverse services.

\end{abstract}


\renewcommand{\thefootnote}{}
\footnotetext{This research was supported in part by the MSIT (Ministry of Science and ICT), Korea, under the ICT Creative Consilience Program (IITP-2020-0-01821) supervised by the IITP (Institute for ICT Planning \& Evaluation). \emph{(Corresponding author: Dong In Kim.)}}

\footnotetext{Gaosheng Zhao and Dong In Kim are with the Department of Electrical and Computer Engineering, Sungkyunkwan University, Suwon 16419, South Korea (e-mail: gaosheng@skku.edu, dongin@skku.edu).}
\IEEEpeerreviewmaketitle



\section{Introduction}

\IEEEPARstart{R}{}ecently, driven by emerging technologies such as multi-sensory immersive experiences, holographic displays, virtual reality (VR), augmented reality (AR), and blockchain, future communication networks are encountering unprecedented challenges \cite{Tataria20216G,Tang20223The}. On the one hand, at the infrastructure level, these networks must support large-scale device connectivity, high resilience, and scalable topologies. On the other hand, at the application level, they are expected to deliver ultra-low latency, mobility support, and seamless integration with emerging services such as the metaverse.

{In this context, digital twin (DT) technology, which has achieved notable success in aerospace and industrial domains, presents a promising direction for next-generation networks \cite{Tao2019digital}. The earliest practical example of DT can be traced to NASA’s Apollo 13 mission in 1970, where ground-based simulators replicated spacecraft systems to assist in incident diagnosis \cite{allen2021digital}. While such digital representations had long existed, the term ``digital twin'' was formally introduced by Dr. Michael Grieves in 2002 to describe a system involving bidirectional data flow between a physical object and its virtual replica \cite{grieves2023digital}. A complete DT system is now recognized as comprising a physical entity, its virtual model, and a real-time data synchronization mechanism \cite{liu2021review}.}

{Therefore, by constructing a twin network on top of the physical communication network, with heterogeneous communication links serving as the bidirectional data exchange mechanism, a DT system of communication network can be established. This architecture further enables intelligent decision-making processes to occur within the DT layer and provides the foundation for advanced services such as the mobile metaverse. However, unlike a single spacecraft or industrial scenarios, communication networks are inherently distributed, highly dynamic, and span large geographic areas with diverse endpoints \cite{Khan2022digital}. The centralized DT architecture relying solely on cloud processing introduces significant latency and scalability bottlenecks, particularly for edge and local devices. A more suitable approach is to develop a multi-layer distributed DT system that reflects the decentralized nature of modern communication environments.}

Following this approach, this paper proposes a multi-layer DT system based on the cloud–edge–local architecture envisioned for 6G networks, as illustrated in Fig. \ref{fig:1}. The proposed system introduces three functional types of DTs, namely local DT, edge DT, and cloud DT, which interact across layers through data, communication, and service coordination modules to bridge the physical and virtual spaces. The main contributions of this paper are summarized as follows:

\begin{figure*}[t]
	\centering
	\includegraphics[width=0.9\linewidth]{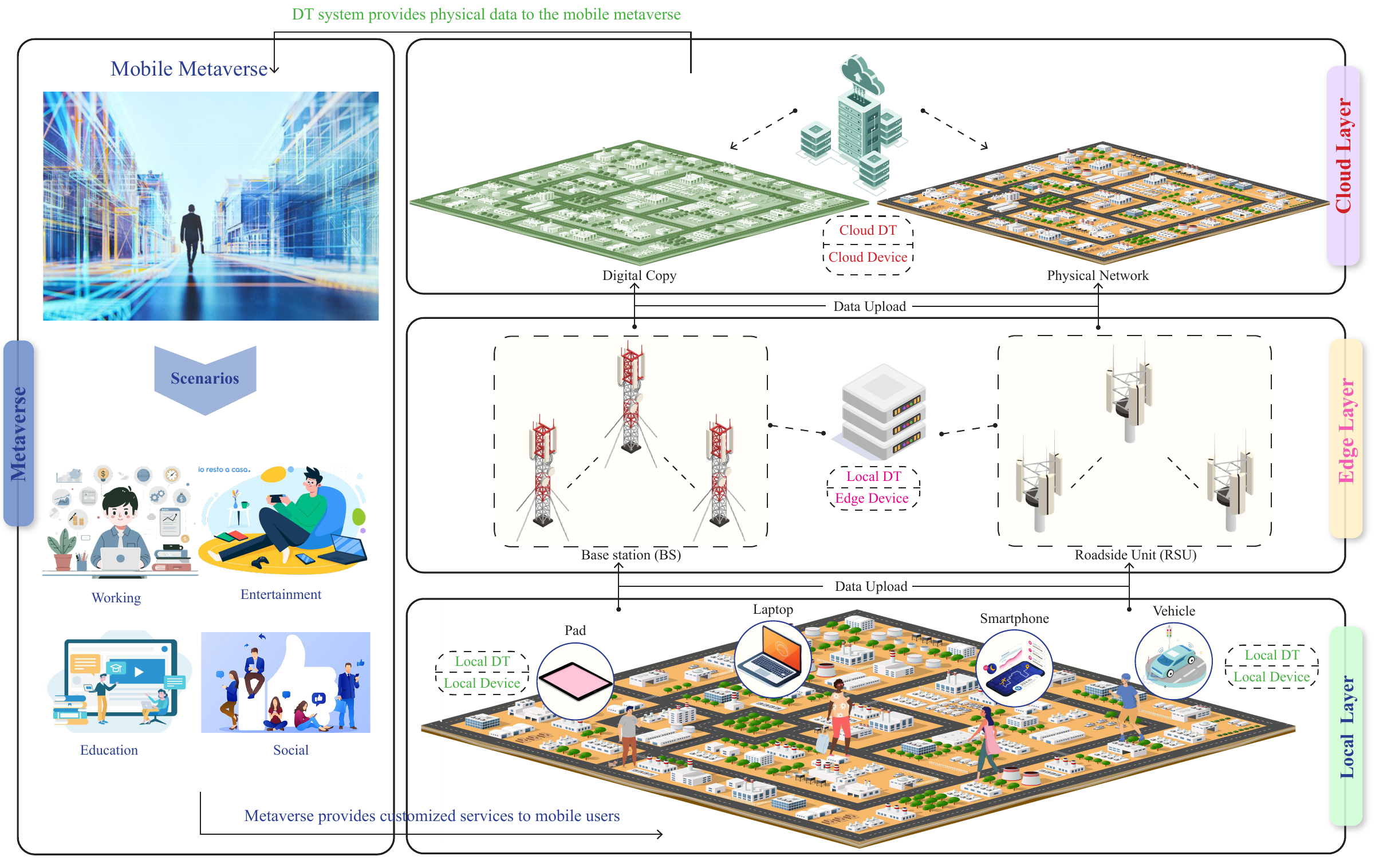}
	\caption{The relationship between the proposed DT system and mobile metaverse.}
	\label{fig:1}
\end{figure*}

\begin{itemize}
	\item \textbf{Layered DT Construction}: According to the physical and application requirements of future communication network, we design a multi-layer DT system for the mobile metaverse. This system constructs a layered DT architecture for the communication network, enabling it to address specific demands of terminal users at different DT layers.

	\item \textbf{Comparison with Centralized Solution}: We first theoretically analyze the new characteristics of the multi-layer DT system compared to traditional centralized approaches. Subsequently, we demonstrate its improvements in real-time feedback latency through a detailed analysis.
	
	\item \textbf{Support for Mobile Metaverse}: We examine how multi-layer DT system can support future mobile metaverse. In short, the proposed DT system introduces metaverse functions during model deployment while simultaneously facilitating the integration of the metaverse with other autonomous systems through open interfaces.

	\item \textbf{Potential Research Directions}: We highlight several intriguing research directions in local, edge and cloud DT to further promote the deployment of the proposed DT system.
\end{itemize}

The remainder of this paper is organized as follows. Section II reviews the centralized DT. Section III shows the details about multi-layer DT system. Section IV discusses a comparison with centralized DT solution by theoretical and case study. Section V explains its support for mobile metaverse. Section VI highlights potential research directions, and finally, the paper concludes with Section VII.

\section{DT Review}

Currently, centralized DTs dominate as the absolute mainstream in DT research and practice. Many DT products, such as Microsoft Azure Digital Twins and Amazon Web Services IoT TwinMaker, are designed to create centralized, unified, and high-fidelity digital representations of systems, workshops, devices, or networks, consolidating all information in one place for human interaction and intervention \cite{HAKIRI2024110350}. For example, in manufacturing system, \emph{Liu et al.}\cite{liu2023digital} utilized DT technology to enable digital collaboration among manufacturing units, dynamically optimizing the overall manufacturing process.  

Owing to the significant impact of DT in the industrial field, researchers are also exploring the integration of DT functionalities directly into the next-generation 6G communication systems. This includes creating digital representations of wireless network assets and components, such as core network, network typologies, radio access network, and user behaviors, within a unified platform to enable dynamic and intelligent management \cite{computers11050067,Huang2024when}.

However, due to the massive scale of communication network and mobility of wireless terminals, it is challenging to build a centralized unified platform capable of representing all network information. Additionally, the significant latency caused by heterogeneity between network terminals and the centralized platform further complicates this approach. These factors not only hinder the application of DT in this field but also make it even more difficult to provide the necessary support for future mobile metaverse.

In this paper, we develop a decentralized multi-layer DT system. By distributing local DT, edge DT, and cloud DT across different network locations, the proposed system embeds DT functionalities natively into the communication architecture to support different digital services in three layers. Furthermore, the multi-layer DT system will supply metaverse data, pre-trained models, and interfaces with other autonomous systems, enabling a wider range of future mobile metaverse.

\section{Multi-layer DT System}

{Inspired by the structure of the human nervous system, where the peripheral system collects stimuli and the central system generates decisions at multiple levels, the proposed multi-layer DT system follows a similar hierarchical model. Just as the spinal cord can autonomously trigger reflexes while the brain coordinates complex actions, lower-layer DTs (e.g., local DTs) respond to time-critical data with fast, localized decisions, while upper-layer DTs (e.g., edge and cloud DTs) perform global coordination across terminals. The proposed multi-layer DT system thus adopts this hierarchical structure to support distributed intelligence and responsiveness and encourages a bottom-up DT construction—from terminals to edge to cloud—enabling essential DT services such as real-time feedback and intelligent agents.

\begin{figure*}
	\centering
	\includegraphics[width=\linewidth]{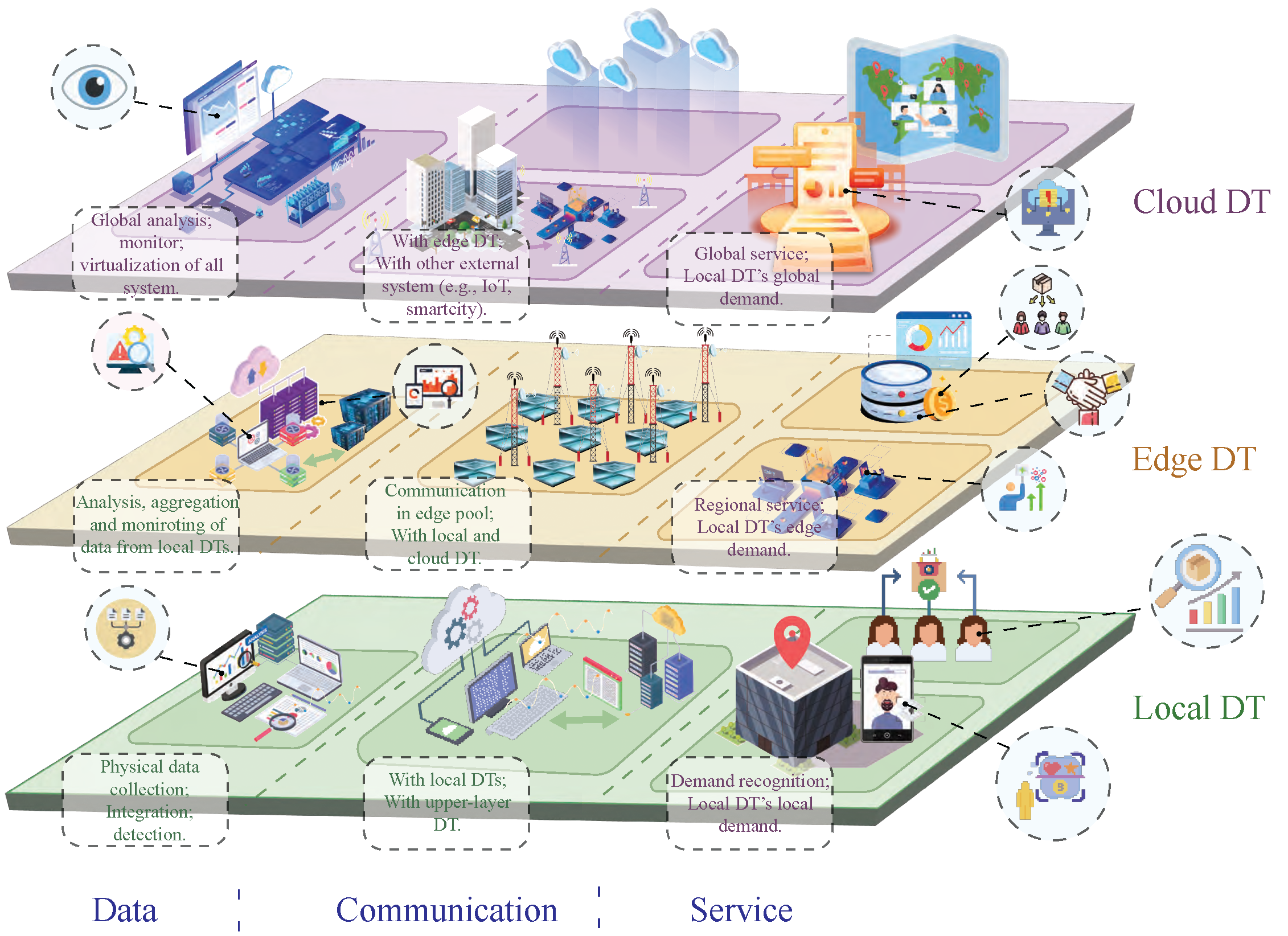}
	\caption{The architecture of the proposed multi-layer DT system.}
	\label{fig:2}
\end{figure*}

To realize this architecture in practice, each layer of the DT system is designed with a modular structure comprising three key components: data, service, and communication modules. The following subsections provide a layer-by-layer breakdown, starting from the terminal-level local DT, progressing to the regional edge DT, and culminating in the cloud DT that provides global coordination, as shown as Fig. \ref{fig:2}.}

\subsection{Local DT}

{At the bottom of the architecture, the local DT is a digital module deployed on the user's terminal.} It is primarily responsible for data collection and processing, partial decision-making, and communication with upper layers. These functions enable the local DT to monitor the physical environment and status in real-time, while also executing low-latency decisions based on localized data. Moreover, local DTs serve as the entry point for DT construction, initiating data-driven interactions upward through the system hierarchy.

To support these roles, the local DT is implemented using a modular design, consisting of the following three core components:

\subsubsection{Data Module}

The data module collects internal, external, and behavioral data to provide a comprehensive view of device performance and user behavior. Specifically, internal data (e.g., CPU usage, memory utilization) reflects the operational state of the device, external data (e.g., environmental sensors, radar inputs) captures ambient conditions, and behavioral data (e.g., application usage patterns) helps infer user intent. This module also performs data cleaning, denoising, standardization, and compression to enhance storage and transmission efficiency.

\subsubsection{Service Module}

The service module is responsible for managing various demands generated by the local entity. In our system, these demands are categorized into local, edge, and cloud levels according to their required resources and latency sensitivity \cite{Zhao2023trine}. When a demand is generated, the service module first determines whether it can be processed locally or needs to be escalated. For local demands, the module invokes lightweight pre-trained models and schedules them based on a priority system, ensuring that time-critical requests (e.g., safety alerts) are served with appropriate urgency.

\subsubsection{Communication Module}

The communication module governs interactions between the local DT and the surrounding DT network. It enables upward communication with the edge DT by transmitting status updates and demand packets, and facilitates downward feedback reception. In the absence of edge infrastructure, the local DT also supports peer-to-peer communication with neighboring local DTs, allowing collaborative decision-making in dynamic or infrastructure-less environments.

\subsection{Edge DT}

{Building upon the information aggregated from multiple local DTs, the edge DT resides at edge nodes (e.g., base stations or roadside units) and acts as the regional control hub within the DT system.} Its main objective is to coordinate distributed entities within a local coverage area, and to optimize regional services under constrained resources.

A key feature of the edge DT is its dynamic pooling structure, which allows digital agents from local DTs to flexibly join or leave the edge DT environment. This design enables real-time adaptation to device mobility and supports scalable coordination across diverse terminals.

Like the local DT, the edge DT is structured into three major modules:

\subsubsection{Data Module}

This module handles the aggregation, filtering, and analysis of data streams uploaded from local DTs. It consolidates individual data sources into a regional dataset, which supports edge-level inference and service orchestration. In addition to data fusion, it performs anomaly detection to flag irregularities that may indicate systemic failures or security threats. Selected data are then forwarded to the cloud DT for further analysis or long-term storage.

\subsubsection{Service Module}

At this layer, the service module supports both regional-layer services (e.g., multi-agent coordination) and intermediate routing of local or cloud demands. Specifically, it (a) regulates and prioritizes edge-layer services in response to high load or unexpected events, (b) adjusts resource allocation across local DTs to maintain fairness and reliability, and (c) receives model updates from the cloud DT. When demands exceed the processing capabilities of the edge layer, they are escalated to the cloud DT.

\subsubsection{Communication Module}

The communication module at the edge layer handles three types of interactions: (a) intra-layer communication with local DTs via their digital agents; (b) inter-agent communication among terminals' agents in the pooling structure to support collaborative services \cite{luan2021paradigm}; and (c) inter-layer communication with cloud DTs for receiving global updates and uploading regional summaries. 

\subsection{Cloud DT}

{At the top of the hierarchy, the cloud DT performs global-level processing and long-term optimization.} It collects regional data from multiple edge DTs, constructs a global system state, and updates global models for network-wide deployment.

While the local and edge DTs focus on short-term responsiveness, the cloud DT emphasizes long-term analytic, macro-level coordination, and external interfacing with cloud services or public infrastructure.

\subsubsection{Data Module}

The data module in the cloud DT aggregates information from multiple edge nodes. It enables global-scale anomaly detection, long-horizon forecasting, and system-wide pattern recognition. These insights can be used to train or refine models for the lower DTs.

\subsubsection{Service Module}

The cloud DT oversees both inter-edge coordination and strategic deployment of resources for applications such as large-scale traffic control, or cross-domain optimization. It dynamically allocates cloud-side resources to serve escalated demands and coordinates actions across multiple edge domains. For local-layer services, cloud DT allocates resources based on the cloud demands of local DTs, while also facilitating top-down communication across cloud, edge, and terminals. 

\subsubsection{Communication Module}

Finally, the communication module ensures reliable downstream propagation of cloud-generated models, strategies, and global policies. It also supports integration with external platforms (e.g., metaverse, blockchain, Internet-of-Things platforms), allowing the multi-layer DT system to adapt to evolving service demands and heterogeneous infrastructures.

{
\section{Comparison with Centralized DT}

To demonstrate the difference of the proposed multi-layer DT system, this section presents a comparative analysis against traditional centralized DT architectures. We examine the differences in design philosophy, associated deployment challenges, and performance under simple communication network.

\begin{table*}
	
	\caption*{TABLE I: Comparison Between Centralized and Multi-layer DT Architectures.}
	\label{tab:1}
	\centering
	\renewcommand{\arraystretch}{1.35}
	\begin{tabular}{|>{\centering\arraybackslash}p{4cm}|
			>{\centering\arraybackslash}p{6.1cm}|
			>{\centering\arraybackslash}p{6.1cm}|}
		\hline
		\textbf{Dimension} & \textbf{Centralized DT} & \textbf{Multi-layer DT} \\
		\hline
		\multicolumn{3}{|c|}{\textbf{System Architecture}} \\
		\hline
		Control Scope & Single DT instance at the cloud level with full control & Hierarchical DTs across local, edge, and cloud with coordinated roles \\
		\hline
		Twinning Pattern & Only vertical twinning (per entity, no layer split) & Supports both horizontal (per layer) and vertical (cross-layer entity) twinning \\
		\hline
		Cross-Device Collaboration & Relies on cloud to mediate all device interaction & Edge DT pooling enables agent-based terminal coordination \\
		\hline
		Scalability & Limited by cloud bandwidth and central processing & Supports adaptive scaling by deploying more edge/local DTs \\
		\hline
		
		\multicolumn{3}{|c|}{\textbf{Performance and Adaptability}} \\
		\hline
		Latency & High due to remote cloud access for all services & Reduced latency via local/edge processing and demand-aware routing \\
		\hline
		Mobility Handling & Poor adaptability under mobility or intermittent link & Local and edge DTs adapt to movement and temporal variation \\
		\hline
		Real-Time Feedback & Typically delayed by cloud inference loop & On-device or edge-layer response to satisfy time-sensitive needs \\
		\hline
		Resource Allocation & Centralized allocation may lead to overload or underuse & Hierarchical allocation based on capacity and context \\
		\hline
		
		\multicolumn{3}{|c|}{\textbf{Deployment and Maintenance Challenges}} \\
		\hline
		Infrastructure Overhead & Easier to deploy with a single-point architecture & Requires distributed deployment, monitoring, and version control \\
		\hline
		Synchronization & Centralized consistency is trivial to maintain & Cross-layer synchronization and data coherence are complex under mobility \\
		\hline
		System Complexity & Lower software/agent lifecycle complexity & High complexity in agent orchestration and multiscale coordination \\
		\hline
	\end{tabular}
\end{table*}

\subsection{Architectural and Deployment Insights}

The architectural distinction between centralized and multi-layer DT systems lies in their control granularity and deployment logic. The centralized DT architecture aggregates data processing, and decision-making into a singular cloud-level twin upon the cloud server. While such architecture simplifies management, it suffers from centralized device bottlenecks, limited scalability, and high response latency, especially in mobile, multi-layer or multi-terminal scenarios. In contrast, the proposed multi-layer DT system adopts a hierarchical structure inspired by the human nervous system. Specifically, local DTs act on immediate environmental feedback, edge DTs coordinate among nearby devices, and cloud DTs offer global analytic and optimization. This layered hierarchy enables both horizontal twinning (multiple DTs per layer) and vertical twinning (single-entity DTs spanning layers), promoting fine-grained control and improved scalability.

In terms of system performance and adaptability, the proposed architecture demonstrates clear advantages. By enabling localized decisions, the system significantly reduces latency for time-sensitive demands and enhances adaptability in mobile 6G scenarios. Moreover, edge pools through digital agents facilitates cross-terminal interaction with minimal backhaul communication. Resource allocation can be dynamically adapted according to the needs of entities—local DTs handle lightweight and tight deadline processing, edge DTs address regional demands, and cloud DTs manage long-term optimization. These functional decompositions alleviate cloud overload to enable better overall efficiency.

Nevertheless, these benefits are accompanied by added complexity in deployment and maintenance. Multi-layer DT system requires multi-site deployment, inter-layer synchronization, and continuous management of distributed digital agents. Fortunately, future communication infrastructures such as 6G offer built-in features including ultra-reliable low-latency communication, AI-native architecture, and deterministic networking, which can furtherly support tight synchronization, adaptive data exchange, and intelligent orchestration. Although the distributed nature of multi-layer DTs introduces security concerns at edge and terminal levels, these can be mitigated through secure communication protocols, federated anomaly detection, and hierarchical trust enforcement.

In summary, the evolution from centralized to multi-layer DTs presents a trade-off between enhanced system intelligence and increased deployment complexity. Centralized DTs are suitable for static, resource-rich environments like industrial automation, whereas multi-layer DTs are better suited for dynamic, latency-sensitive, and mobile scenarios such as smart transportation and layered network. The distributed architecture improves responsiveness, mobility support, and service diversity, while requiring more from synchronization, security, and lifecycle management. Overall, multi-layer DTs offer greater advantages for future distributed and mobile networks. The comparative characteristics of both architectures are summarized in Table~I.

\subsection{Case Study: Multi-layer DT Deployment in a Typical Communication Network}

To further validate the proposed architecture, we conduct a case study in a typical hierarchical wireless communication network. The network consists of ten local terminal devices, two edge servers, and one centralized cloud node, respectively equipped with lightweight (1 GFLOPS), medium-scale (20 GFLOPS), and high-performance (500 GFLOPS) computing capabilities. Each local device communicates with the edge layer via wireless links (uplink rate: 50 Mbps; downlink rate: 200 Mbps), while edge-to-cloud transmission is supported by wired optical fiber links (1 Gbps bidirectional bandwidth). We compare two DT deployment strategies and Fig.~\ref{fig:3} illustrates the difference between the two DT architectures in handling multi-layer demands generated by local physical entities: 
\begin{itemize}
	\item \textbf{Centralized DT:} A single global DT is deployed at the cloud server. All terminals transmit raw data to the cloud for modeling and decision-making. Both edge and local nodes act as passive relays, without autonomous intelligence or contextual awareness.
	
	\item \textbf{Multi-layer DT:} DTs are deployed at each layer. Each terminal hosts a lightweight local DT that reacts to immediate physical changes and performs time-sensitive decisions. Edge servers maintain regional DTs to coordinate multiple terminals, perform data aggregation, and enforce local policies. The cloud layer integrates system-wide data and handles long-term analysis, global optimization, and model refinement.
\end{itemize}

\begin{figure*}[t]
	\centering
	\includegraphics[width=\linewidth]{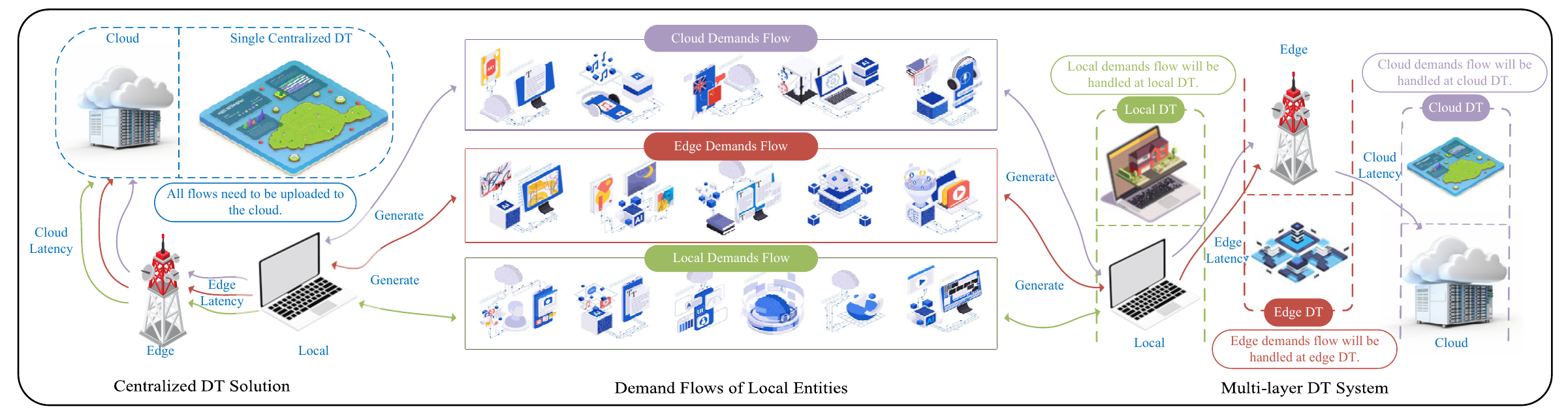}
	\caption{The difference between the two DT architectures in handling multi-layer demands.}
	\label{fig:3}
\end{figure*}

Fig.~\ref{fig:4}(a) compares the end-to-end response latency of centralized and multi-layer DT architectures under a high-mobility scenario, where 5 out of 10 local devices switch their associated edge server every second. This mobility introduces frequent handovers and dynamic routing adjustments. The centralized DT, relying solely on cloud processing, suffers from higher and more volatile latency (870–930 ms) due to limited adaptability to rapid context changes. In contrast, the multi-layer DT system maintains stable low latency (330–360 ms) by leveraging local and edge-level DTs for immediate response and buffering, effectively mitigating the impact of mobility-induced disruptions.

\begin{figure*}
	\centering
	\begin{minipage}{0.48\textwidth}
		\centering
		\includegraphics[width=\linewidth]{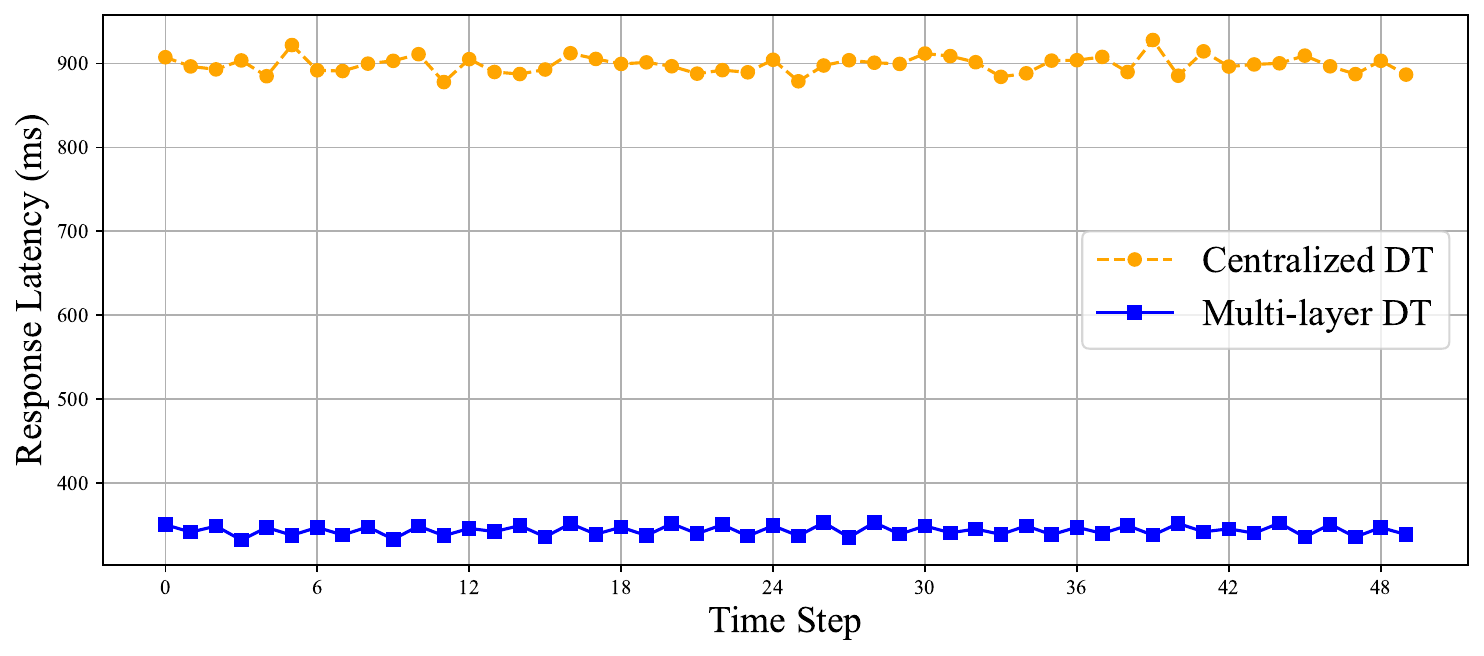}
		\begin{flushleft}
			\centering
			\small (a)
		\end{flushleft}
	\end{minipage}
	\hfill
	\begin{minipage}{0.48\textwidth}
		\centering
		\includegraphics[width=0.82\linewidth]{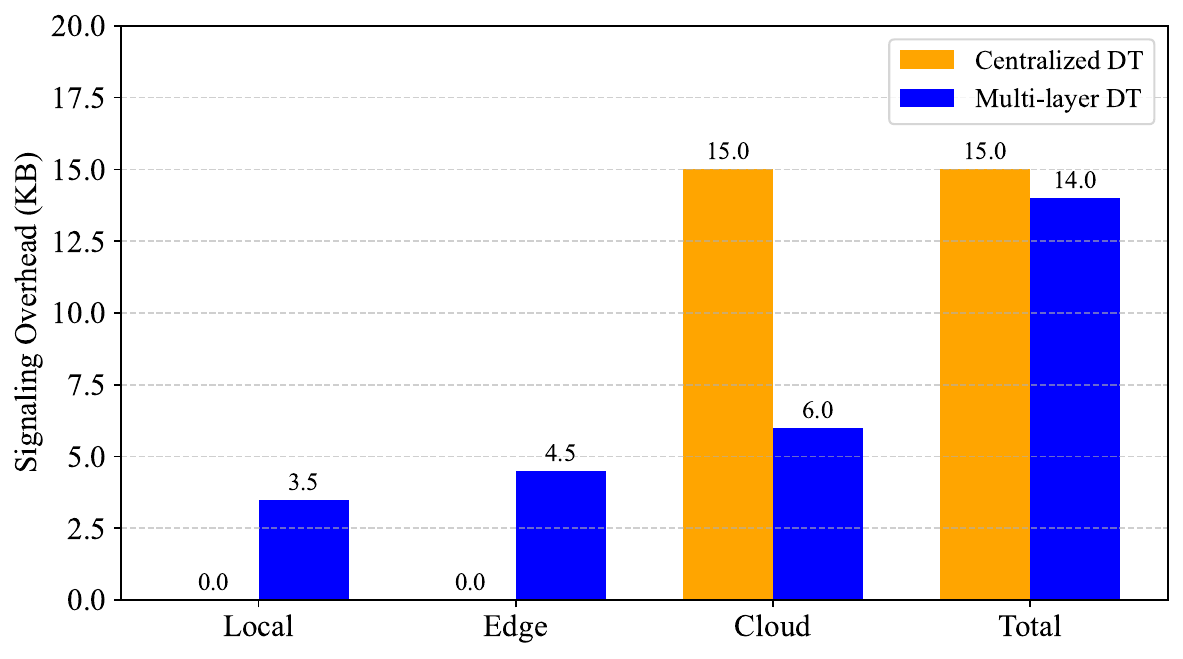}
		\begin{flushleft}
			\centering
			\small (b)
		\end{flushleft}
	\end{minipage}
	\vspace{-0.5em}
	\caption{ Comparison between centralized and multi-layer DT architectures: (a) Response latency in high-mobility scenarios; (b) Signaling overhead distribution across DT layers.}

	\label{fig:4}
\end{figure*}

Fig.~\ref{fig:4}(b) further illustrates signaling overhead distribution across centralized and multi-layer DT architectures under varying levels of demand complexity. While the centralized DT relies entirely on the cloud, resulting in consistently high signaling overhead (15~KB per demand) \cite{Tao2024Wire,Yu2025optimal}, the multi-layer DT enables hierarchical offloading, with demands dynamically processed at local, edge, or cloud layers. Local DTs typically generate lightweight signaling traffic composed of status summaries, semantic results, and protocol headers, totaling approximately 3.5~KB per demand. Edge DTs, supporting mid-level coordination and regional inference, contribute moderate signaling overhead around 4.5~KB. Only complex demands escalate to the cloud layer, incurring higher signaling loads.  By enabling early-stage decision-making and reducing redundant transmissions, multi-layer DT system retains comparable signaling overhead while achieving superior adaptability and scalability in dynamic environments.

}

\section{Support for Future Mobile Metaverse}

To ensure seamless and efficient user experiences in mobile environments, future mobile metaverse also develops its own applications and services according to 6G cloud-edge-terminal architecture. Specifically, the cloud generates optimized models, which are distributed to the edge and terminal layers, ensuring the entire metaverse is updated with the latest intelligence and providing consistent, high-quality user experiences regardless of mobility. Edge nodes process user data closer to its source, reducing communication latency between terminal and cloud layers, and improving response speed. The terminal layer serves as the entry point for user interaction with the metaverse, typically running on mobile devices such as smartphones, tablets, and wearables. 

\begin{figure*}
	\centering
	\includegraphics[width=\linewidth]{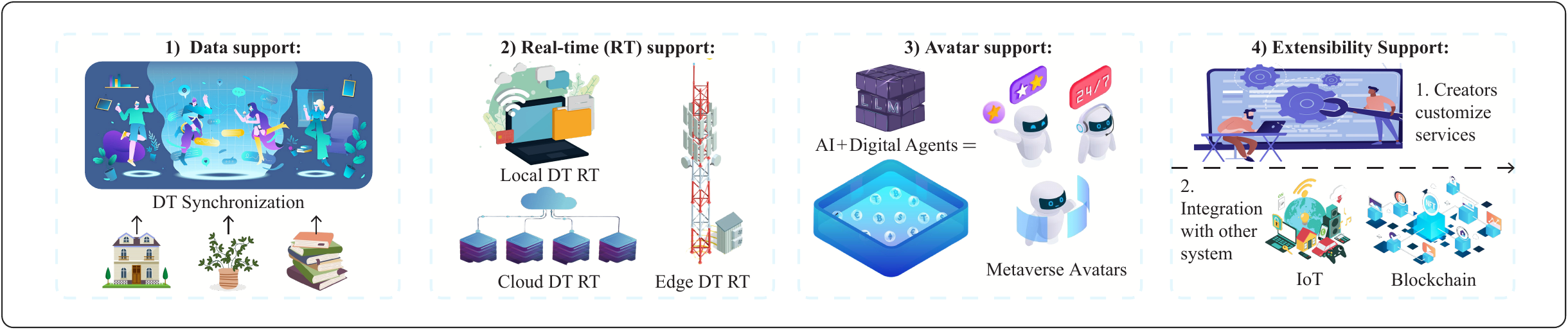}
	\caption{Support for future mobile metaverse.}
	\label{fig:5}
\end{figure*}

The multi-layer DT system, natively embedded in the 6G network, exhibits distinct advantages in supporting the future mobile metaverse. Through open cloud DT interfaces, the system enables seamless configuration of metaverse functions across cloud, edge, and terminal layers. Due to its distributed architecture and various services, as described in Fig. \ref{fig:5}, the system will effectively support future mobile metaverse in terms of data, real time, digital avatar and extensibility.

\begin{itemize}
	\item \textbf{Data Support}: Multi-layer DT system can collect and transmit data from the physical world to the metaverse, allowing users to experience a highly immersive environment that spans scenarios. This data support brings the virtual environment closer to reality.

	\item \textbf{Real-time Support}: Multi-layer DT system distributes real-time feedback for various metaverse services across different layers and mobility-aware. The local DT handles terminal-layer responses by processing local needs using cached and collected data and detect mirco-movements of users. The edge DT extends to regional macro-mobility of terminal devices, enabling edge-level metaverse service responses. Finally, the cloud DT offers global services and model optimization support, serving as the final link in the real-time response chain.

	\item \textbf{Avatar Support}: Digital agents within the system can integrate with large language models (LLMs) and other advanced technologies to serve as virtual avatars for human users in the metaverse. These avatars operate primarily in the edge layer, enabling both intra-twin and inter-twin interactions. They also coordinate demands and resources across the local and cloud layers. 

	\item \textbf{Extensibility Support}:
	Scalability is reflected in two key aspects. First, through the cloud DT interface, creators can customize services by directly embedding metaverse functionalities into DT modules. Second, this interface facilitates seamless integration of the metaverse with other emerging systems and technologies. For example, by connecting with Internet of Things (IoT), the metaverse gains interactive contextual awareness. Additionally, integration with blockchain technology ensures secure management and transparent transactions of metaverse assets.
\end{itemize}

\section{Potential Research Directions}

In this section, we outline several future research directions for DTs located at different layers within the system, including local, edge, and cloud DTs. 

\subsection{Local DT}
At the local layer, the primary goal is to integrate the existing scattered digital functionalities of local entities into a unified system. This involves comprehensively consolidating independently managed digital assets, data, and functionalities and transitioning them into a unified DT model.


Key directions include optimizing hardware and software resource allocation to minimize latency for real-time interactions on terminal devices and developing advanced models that leverage device-interactive AI for immediate requirements feedback. Additionally, exploring differentiated modes will allow local DTs to adapt to diverse and dynamic user demands, such as work, entertainment, and varying activity patterns in hotspot and non-hotspot areas.

\subsection{Edge DT}
Edge DT is the most critical and structurally complex component of the multi-layer DT system, with its primary role being the connection between local DTs and cloud DT. On the one hand, edge DTs are tasked with constructing a DT pool that supports the coexistence of multiple digital agents, while serving as a platform for coordinating regional resources to meet the edge demands of local entities. On the other hand, as a bridge to cloud DT, edge DTs receive updated models from the cloud, distributing them to local entities through an effective update mechanism. In addition, they transmit data and requests from local entities to the cloud.

For edge DT, feasible future research directions include designing intelligent algorithms to optimize the allocation of edge computing, storage, and communication resources to meet the demands of physical domain; developing LLM-based intelligent agent models to enable the digital representation of metaverse users at the edge layer; investigating the decomposition of complex metaverse functions into independent, operable DT modules and integrating these modules into edge DTs. Furthermore, allowing the edge layer serves as a platform for coordinating local entity demands and cloud resources.

\subsection{Cloud DT}
Cloud DT acts as a model updater, resource repository, and open interface hub at the top of the DT system. It provides a comprehensive array of interfaces, enabling internal compatibility with emerging technological applications and efficient external integration with other autonomous systems. For instance, by incorporating metaverse functionalities into its models, the DT system can address the metaverse's data dependency on physical environments and deliver cross-sensory real-time feedback to metaverse users.

The future of cloud DT research encompasses several key directions. First, developing standardized frameworks and automated tools to streamline the generation, deployment, operation, and updating of DTs across cloud-edge-local layers. Second, enhancing integration with other autonomous systems (e.g., smart cities, IoT networks, mobile metaverse) and emerging technologies (e.g., blockchain, AI) through standardized interfaces and protocols. Third, strengthening cloud-edge-terminal collaboration to ensure consistent, high-quality user experiences across devices and locations.

{
\subsection{Cross-layer DT Collaboration}

Cross-layer DT collaboration refers to the cooperative operation of DTs distributed across the local, edge, and cloud layers. Each DT layer is tailored to a specific scope of control and responsiveness. However, without deliberate collaboration strategies, this layered structure may suffer from state inconsistency, duplicated decisions, or inefficient task handovers. These issues are especially critical in high-mobility and densely connected environments envisioned in future 6G networks.

To support robust collaboration, future research should explore mechanisms such as distributed DT orchestration, dynamic role assignment, and inter-layer abstraction interfaces. Additionally, techniques like hierarchical model synchronization, context-aware information fusion, and priority-based update scheduling can enhance cooperation efficiency. A well-coordinated cross-layer DT framework not only ensures consistency and continuity across layers, but also improves system scalability, responsiveness, and resource utilization under dynamic and heterogeneous conditions.
}

\section{Conclusion}

In this paper, we present a device-oriented multi-layer DT system designed for the future user-oriented mobile metaverse and communication networks. The system operates on two key fronts: analyzing physical data and delivering real-time services through its hierarchical architecture, while supporting open interfaces, digital avatars, and addressing the challenges of mobility and scalability. In addition, we also emphasize the advantages of this solution over traditional centralized DT system and propose potential research directions. By leveraging its layered structure, edge DT pools with digital avatars, and cloud-based model updates, the proposed DT system will lay a scalable foundation for immersive, responsive, and user-centric mobile metaverse.

\bibliographystyle{IEEETran}
\bibliography{ref}

\section*{Biographies}
\vspace{-8mm}
\begin{IEEEbiographynophoto}{Gaosheng Zhao}(gaosheng@skku.edu)
	is currently pursuing the Ph.D. degree with the Department of Electrical and Computer Engineering, Sungkyunkwan University, Suwon, South Korea. His research interests include digital twins, metaverse, and future communication networks.
\end{IEEEbiographynophoto}
\vspace{-10mm}
\begin{IEEEbiographynophoto}{Dong In Kim}(dongin@skku.edu)
	received the Ph.D. degree in Electrical Engineering from the University of Southern California, Los Angeles, CA, USA, in 1990. He is a Distinguished Professor with the Department of Electrical and Computer Engineering, Sungkyunkwan University, Suwon, South Korea. His research interests include the Internet of Things, wireless power transfer, and connected intelligence.
\end{IEEEbiographynophoto}
\vfill

\end{CJK}

\end{document}